# Prismal view of ethics

Sarah Isufi[1], Kristijan Poje[2], Igor Vukobratovic[3], Mario Brcic[2]

[1] ForHumanity

[2] University of Zagreb, Faculty of Electrical Engineering and Computing

[3] Independent researcher

sarahisufi@forhumanity.center, kristijan.poje@fer.hr, vukobratovic.igor.1988@gmail.com, mario.brcic@fer.hr

*Sometimes one needs look back to find out where she is going.*

**Abstract**

We shall have a hard look at ethics and try to extract insights in the form of abstract properties that might become tools. We want to connect ethics to games, talk about the performance of ethics, introduce curiosity into the interplay between competing and coordinating in well-performing ethics, and offer a view of possible developments that could unify increasing aggregates of entities. All this is under a long shadow cast by computational complexity that is quite negative about games. This analysis is the first step toward finding modeling aspects that might be used in AI ethics for integrating modern AI systems into human society.

## 1 Introduction

Life is rich with challenges, decision-making, and questions that we pose to ourselves. Decision-making takes place within a context whose characteristics we will refer to as ***The Setting***. *Ethics* is a discipline concerned with moral values (good and bad) and norms (right and wrong). Norms define standards of acceptable behavior by groups. Specific ethical systems, through their norms (computable conventions), constrain and partially solve the problem of life. The importance of ethics for society is paramount as no social group can stay cohesive and in existence if there are no constraints on the behavior of individuals. For example, frequent, reasonless escalations and attacks with killing or injuring others would dissolve any group. Purzycki et al., 2022 refer to morality as pro- or anti-social norms with direct benefit or cost to others (e.g., theft, murder, generosity, sharing).

Throughout human history, significant technological and cultural advancements have occured in several last millennia. The speed with which these changes arrived was accelerating. However, it was still pedestrian pace compared to the changes coming with greater connectivity (internet), stronger computation (Moore's and descendant laws), cognitively powerful non-human entities (artificial intelligence), and many other disruptive technologies made possible by those. Strong computation and powerful algorithms introduce powerful, flexible, and fast-changing entities into society while the connectivity diffuses the effects of their actions to all corners of the world. All social groups will become paired with these artificial entities, and social adaptation and integration will, due to the speed of changes, be tested as never before. Technology that is the source of difficulties in the first place can, through its dual use, also be used to help alleviate the problem. Wittgenstein suggested a pragmatic view on the development of language of through language-games (Wittgenstein, 1953). We wish to pursue a similar line of thought with ethics and investigate its properties from the computational perspective. We shall tease out different properties that might help in modeling, simulating, and potentially innovating ethical systems that will circumvent issues and deliver us to the good side of future history.

Cooperation was a topic of thorough research conducted and surveyed from the perspective of social (Bowles & Gintis, 2013; Henrich & Muthukrishna, 2021) and natural sciences (Jusup et al., 2022). The former have approached the problem from the top through empirical studies on people. They face interpretation problems because scarce results under-constrain the studied complex setting and leave a multitude of plausible interpretations. Social physicists have approached the problem bottom-up by researching the evolution of cooperation in a simplified utilitarian setting of social dilemma games with low strategic complexity. This narrow focus has enabled them to establish a richness of rigorous conclusions. However, their applicability to realistic cases is quite limited for several reasons. Simplifying assumptions that need to be made for computational reasons also limit the transfer of results to other situations. Cooperation in social dilemma games is only one form of a more general class of moral behavior (Capraro & Perc, 2018). Preferences of agents in some situations cannot be entirely explained just by the monetary outcomes of games, but following own personal norms can offer a better explanation (Capraro & Perc, 2021). Moreover, Bowles, 2016 claims that incentives and social preferences are not separable, and the former affect the latter. Additionally, Broome, 1999 criticizes approaches that assume a single objective that affects each agent's decision-making. Although cooperation is much better understood than before, there are no conclusive answers to the most important questions about cooperation and ethics.

Design for values (value-sensitive and ethically-aligned) is important application of ethics that calls for responsible innovation in the face of accelerating progress which strains the existing social fabric (Helbing, 2021; Hoven et al., 2019). We hold that with the increasing complexity of technology, we are hitting the limits of inference, such as unverifiability and limits to explainability (Brcic & Yampolskiy, 2021), that make that well-intentioned proposition long-term infeasible in the current form due to cognitively superior agents with which value alignment is still a widely open problem.

In this paper, we summed up our contributions as follows:
- We offer synthesis and elaboration of work in different fields related to investigating cooperation and ethics.
- We pivot from the existing practice by focusing on ethics as the first-class mechanism, teasing out its general properties to provide common ground for future interdisciplinary investigations. Additionally, we enrich the description with a computational perspective that relates to computational efficiency. The research in social physics has been narrowly focused and sometimes off-mark by focusing on problems that do not possess these properties. It aimed to show conditions and mechanisms under which cooperation emerges from unbiased and simplified non-cooperative agents. Such generality is a strong a priori requirement, and the found conditions may not even be aligned with our current situation. On the other hand, we accept and carefully describe the current position where humans have significant prosocial bias. The latter position has a much smaller scope.
- We argue for more intentional moral innovation to prepare for coexistence with cognitively superior agents. Current ethics that has so far emerged collaterally has some deficient properties that make value alignment with advanced technologies even more challenging. We can even use technology as a tool for meet-in-the-middle approaches to value alignment. Based on computational complexity considerations, we provide a few pointers as to how this can be made.

Section 2 deals with the main purpose of ethics and its role in group performance. Section 3 explains the ethical dynamics and underlying factors causing ethics to evolve. The multicriteriality of The Setting is further elaborated in Section 4. Furthermore, Section 5 elaborates on the uncertainty of ethics and ethical behavior. The idea of ethics emerging in cultural evolution collaterally is further investigated in Section 6. When considering modeling ethics through game theory and multi-agent systems, Section 7 gives examples of games and modeling choices. Section 8 deals with ethics and its importance in group coordination. Furthermore, the human values in group coordination are elaborated in Section 9. Evolutionary game theory as a modeling basis for ethics is described in Section 10. Social norms as behavioral patterns making up ethics are discussed in Section 11. In Section 12, we elaborate on the consistency of ethics with two mechanisms yielding inconsistency: norm confusion and faulty execution. Finally, Section 13 elaborates on cooperation vs. competition in groups. Conclusions are drawn in Section 14.

# 2     Main purpose of ethics?

We argue that the primary purpose of ethics is achieving better group performance. Coordination is an essential aspect of ethics since it makes collectives more progressive under the cap of available resources to bring about a better outcome for the group. Cooperative societies with a clear division of labor progressed faster because the members were united under the same goal (Becker & Murphy, 1992; Ricardo, 1817).

The cost-effectiveness of cooperation at all stages of social development is an essential item and a prerequisite for determining a posteriori ethics and moral rules. Despite the complexity of moral imperatives in past and present ethics, many have a visible discourse about cooperation within the community. From the slave-owning societies of Greece to South America, from monarchies to republics, there is a rule of respect and cooperation with one's equals. The difference is in the definition of equality and which social, age and class groups fall into that definition, and which are outside it. Thus, cooperation is a plausible precondition for the emergence of ethics per se, no matter how it developed later, whether it included a larger or smaller group of people, the whole community, or just a selected few. The advancements of societies are a by-product of individual satisfaction, which comes from social evolution. In order to maximize the social evolution, freedom of individuals is required because only then the selection process has enough variability to maximize the social fitness. Consequently, the freedom and struggle for survival yield altruism which, next to the cooperation, serves as the backbone of every prosperous society (Dawkins, 1976).

The main reason for increased altruism in society is the selection process, illustrated in the following example. Parents who are not altruistic toward their children will have children with a lower survival rate. Over time, the altruistic population will increase in number, and individuals without those traits will decrease in number (Thompson, 1999). This kin-based altruism (Dawkins, 1976) has a limited range of effects. Another mechanism for cooperation is reciprocity which appears in repeated interactions (Axelrod, 1984). It somewhat increases the range of effects, but is sustainable in dyads and it tends to collapse in larger groups (Boyd & Richerson, 1988; Henrich & Muthukrishna, 2021). Kin-based altruism and long-term interactions are mechanisms through which natural selection on genes can produce cooperation (Lehmann & Keller, 2006) but they are not sufficient to explain the high-level of cooperation observed in humans. Cultural products such as social norms and institutions maintained by mechanisms related to reputation, signaling, and punishment form longer-term cooperations within much larger groups and under a broader range of conditions (Henrich & Muthukrishna, 2021; Wu et al., 2021).

# 3     Ethical dynamics

It has already been mentioned how normative ethics emerges due to social dynamics and cultural trade-offs. Simply put, ethics limits the abilities of person A, so that person A could not harm person B and vice versa. The trade-off is the willing acceptance of a restriction of action to increase welfare for all the factors involved and the general social structure. Therefore, this makes ethics prone to change with changing social standards and ultimately uncertain.

The basis of the claim is the existence of a social consensus on whether a rule will be accepted or not (Assaad, 2021). It is evident that ethics is not static, but it changes dynamically in response to the changes in the environment. One needs not to look too far into the past to see remarkable changes in moral systems during the 20th century (Wheeler et al., 2019). Today's growing technological innovation puts people in new situations that need new societal wisdom, for example, artificial intelligence, globalization, the rise of multinationals, and the metaverse.

The nature of human morality is confined to norms and conventions (Lewis, 1969) describing the individual's behavior and posed rules to regulate individuals and groups. However, society's degree to which individual differences are permitted is variable. The tolerance of society to the variability of individual diversity is crucial to maintaining a biological system that adapts to changes in the environment and throughout time. In general, the rules defining the morality of humans have evolved with the ultimate goal of transmitting genes to succeeding generations. In order to support this goal during various historical accidents, climate changes, and different structures of the gene pools, various beliefs and behaviors have been developed (Petrinovich, 1998).

Another essential aspect of the evolution of morals is that cultural bias, and human values are not genetically predetermined, i.e., humans have multiple behavioral potentials. Despite inherited predispositions, humans have the emotional and cognitive abilities to be both selfish and cooperative. Different circumstances and societies cause each individual to find their moral trajectory (Allchin, 2009). Cultural evolution is another driver of the development of human culture. Humans share information via language and media (e.g., music, writings) that enables the distribution of information and resources, thus providing mechanisms for cultural evolution. Another property of humans accelerating the cultural evolution by freeing cultural information from conceptual limits is *metarepresentation*, i.e., thinking about how we think (Distin, 2011). Unlike genetic evolution causing slow changes in societal culture, cultural evolution has a substantially faster rate (Diamond & Diamond, 1997).

# 4     Multicriteriality of The Setting

Utility theory, which is based on improving a single objective, has been criticized in economics due to apparent incommensurability of options in reality (Broome, 1999). Brcic & Yampolskiy, 2021 hypothesize that human decision-making is made in multicriterial space where the mood selects a subset of focal criteria. These focal criteria are heuristically optimized as near as possible to the Pareto front. Non-focal criteria are simultaneously kept within the acceptable bounds. When there are multiagent interactions, we enter the multicriterial aspects of ethics which tend to create ethical dilemmas. The trade-offs can be between the important drives within an individual or between the benefits of the individual and society. These dilemmas cannot be elegantly resolved. We argue that this can be connected to the property that there are many competing criteria on which ethical decisions must be based, as well decisions in which games to take part at specific moment. The trolley problem (Foot, 1967) is one such problem where the criteria of "do no harm" and "reduce suffering" play against each other and cannot flatly be resolved without being wrong against some criteria. Ethical dilemmas constrain the achievement of perfect outcomes so that it is often impossible to respect multiple criteria simultaneously. This means inevitable trade-offs, defined in, e.g., fairness (Kleinberg et al., 2017) and Social Choice Theory (SCT) (Arrow, 1950), must be made where we choose the solution that achieves the maximal possible hypervolume indicator (Brcic & Yampolskiy, 2021).

Consequentialist ethics is prone to dilemmas originating from multicriteriality whereby several criteria need to be traded off in a consequential state. Deontological ethics can use norms to dissolve complicated commonly occurring dilemmas into more straightforward coordination problems (Bicchieri, 2005). However, such systems introduce dilemmas through inconsistencies, as explained in section 12.

# 5     Status of uncertainty in ethics

Ethical behavior is first and foremost practical activity. Namely, epistemological limits (information and cognition) are not held against actors in the case of mistakes and bad outcomes, instead, they are used for discounting the responsibility. Actors often do not possess sufficient information or necessary cognition to achieve omniscient and omnipotent (and yet, still subject to some limits) solutions. Courts recognize the same principle in the majority of legal systems. For example, a person with temporary or permanent reduced cognitive ability will receive a more lenient sentence.   The primary motivation behind this act is that mental impairment caused by mental illness or substance use diminishes the mental capacity to make rational decisions (Bernard & Gibson, 2003).

Another example of ethical uncertainty is caused by insufficient information. Hindsight bias indicates that human post-fact decisions are likely to be affected by knowing the outcome of their actions. This means humans will reconstruct the entire thinking process leading them to an initial decision when they hear the outcome and change their final decision accordingly (Sligo & Stirton, 1998). Hindsight is discounted from responsibility. For this reason, if a surgeon, for example, misinterprets a patient's diagnosis due to latent factors leading a patient to death, he will not be prosecuted. Had he known the actual diagnosis, he would have taken different actions.

# 6     Collateral nature

There are universal moral rules, but there is no unified ethics (Kinnier et al., 2000) as there is a lot of variation between different moral systems in human culture (Awad et al., 2020). Ethics has always awaited us in the world; it was a forward-handoff from a continuous stream of generations to their posterity. Just as the COVID-19 pandemics have demonstrated, new situations call for new solutions. Since new situations, especially significant ones, are inherently random, ethics has so far emerged collaterally, i.e., under no guidance by some human designer. Societies adjust ethics to technical progress, social conditions, and cultural standards. The question is: Can ethics that does not arise collaterally even be created? Can there be a system for predicting ethics or the best possible moral course for society?

When considering the origin of ethics, we argue that it has emerged collaterally but not randomly. Instead, several factors have influenced the development of ethics, including the neurobiological characteristics of each individual and the sociocultural environment in which the individual develops. Moreover, the essential elements determining the development of moral judgment and consequently functioning when resolving dilemmas are derived from cultural characteristics, spirituality, socioeconomic environment, life experiences, and correct neurological functioning (Hanun Rodríguez & Ximénez Camilli, 2018).

Darwin's view on moral theory is based on conscience, i.e., social instinct. A social instinct is how an individual behaves in a group for that group's benefit. Individual behavior will result from adopted human values, influencing every decision made that has consequences on the group. Consequently, the social instinct results from the group's evolution, increasing group fitness. Unlike other social animals, humans have developed intellect that allows reasoning when faced with dilemmas. However, such reasoning is inevitably constrained by social instinct and human values (Darwin, 2008).

# 7 Ethics, multiagency, and games

Game theory is a branch of science that deals with interactions between different actors, which is exactly the level of operation for ethics. **Classical Game Theory (CGT)** is based on rationality and just-in-time computation interleaved with acting with an unrealistic amount of information and compute. CGT enables simple interepisodic learning (memory) on the level of individual. **Evolutionary Game Theory (EGT)** in the classical form is an application of game theory on evolving populations and it does not require rationality. It is a form of evolutionary policy search where genotype completely describes the lifetime behavior (phenotype). Hence, "learning" in EGT is populational and intergenerational. **Multi-Agent Reinforcement Learning (MARL)** (Du & Ding, 2021; Yang & Wang, 2021) is a more modern framework than the previous two. It enables more complex and structured strategic learning on the level of individual during their lifetime. It scales to more complex group dynamics and strategies than CGT and it brings about individual and lifetime learning in comparison to EGT.

It is plausible that ethics has arisen due to evolutionary processes that a game theory can model. Therefore, it can be represented by an evolutionary model containing a representation of the population's state and a dynamic set of laws influencing the state changes over time. Different mechanisms have been used to explain the rise of cooperation, norms, and ethics in societies: kinship altruism, direct reciprocity, indirect reciprocity, network reciprocity, group selection, and many others (Capraro & Perc, 2018; Henrich, 2004; M. A. Nowak, 2006). They have been analyzed from different perspectives including biologists, political scientists, anthropologists, sociologists, social physicists, economists, etc. The following three concepts are crucial for our exposition. **Nash equilibrium** is a strategy profile from which deviation would not be profitable for any player. **Evolutionary Stable Strategy (ESS)** is a refinement of Nash equilibrium which is evolutionary stable. The population adopting it could not be invaded by mutant strategy through natural selection. Finally, **correlated equilibrium** is a generalization of Nash equilibrium that emerges in the presence of a correlation device.

## 7.1 Examples of games

There is a multitude of games used in literature for theoretical analysis (Capraro & Perc, 2021; Jusup et al., 2022) and behavioral experiments (Henrich & Muthukrishna, 2021). Here we give several examples with results obtained on them.

The **Prisoner's Dilemma** (PD) is one of the fundamental problems of game theory that shows remarkable property that can be connected to emergent ethics based on direct reciprocity (Axelrod, 1984). This problem exemplifies pure competition, and it is the most challenging type of environment for cooperation to appear. Namely, in the case of a single-iteration PD game, the maximum benefit comes from selfish play, that is, from betraying a cooperating partner. However, when the problem is changed to a multi-iteration prisoner's dilemma, we can get cooperation between partners as stable and optimal behavior. By the folk theorem, iterated PD has an abundance of Nash equilibria, and which solving process ends up sensitively depends on the specifics of the environment (Jusup et al., 2022; Stewart & Plotkin, 2014) – with both defection/extortion (Press & Dyson, 2012) and generosity being a possible dominant solution (Stewart & Plotkin, 2013).

In reality, we have observed cooperative behavior also in different contexts, such as where neighbors settle disputes in ways that are not achievable between strangers (Ellickson, 1991). In repeated play, selfishness is charged because the teammate has insight into the player's past moves, making it not profitable to be selfish through direct reciprocity. However, nowadays we have tools such as Internet reputations and social media ratings that are publicly available, giving us insight into players' past moves without previously playing games.

The **Stag Hunt** (SH) problem in game theory originated from Rousseau's *Discourse on Inequality* as a prototype of the *social contract*. It describes the trade-off between safety and cooperation to achieve more significant individual gain (Skyrms, 2004). Unlike the PD problem, where an individual's rationality and mutual benefit are conflicted, in the SH problem, the rational decision is nearly a product of beliefs about what the other player will do. If both players decide to employ the same strategy, stag hunting and hare hunting are the best options. However, if one player chooses to hunt stag, he risks the other player will not cooperate. On the other hand, a player choosing to hunt a hare is not faced with such a risk since the other player's actions do not influence his outcome, meaning rational players face a dilemma of mutual benefit and personal risk (Skyrms, 2001).

**Fair division theory** deals with procedures for dividing a bundle of goods among *n* players where each has equal rights to the goods. Comparing which procedure is the most equitable gives a fair insight into popular notions of equity (Crawford, 1989). The modern theories of fair division are used for various purposes such as division of inheritance, divorce settlement, and frequency allocation in electronics. The most common division procedure is *divide and choose*, used for a fair division of continuous resources. Steinhaus describes it in an example of dividing a cake among two people where the first person cuts the cake into two pieces and the second person selects one of the pieces; the first person then receives the remaining piece (Steinhaus, 1948). Such a game is categorized in the field of *mechanism design*, where the setting of the game gives players an incentive to achieve the desired outcome (Myerson, 1989). However, the procedure proposed by Steinhaus does not always yield fairness in a complex scenario setting since a person might behave more greedy to acquire more of the goods he desires. A procedure that is considered *fair* implies the allocation of the goods should be performed in a manner where no person prefers the other person's share (Foley, 1967).

Evolutionary game theory (EGT) shows in several examples, e.g., PD, Hawk/dove, Stag/hare (Dawkins, 1976), the tendency that cooperation is a better approach in the long run (an iterated relational game), while selfishness tends to be better in the single-step (transactional version of the game). These results of repeated games depend on the settings of problems and the utilization of different mechanisms that support the emergence of cooperation (Capraro & Perc, 2018).

## 7.2 Modeling choices

Two main choices are given when representing the population: using continuous or discrete models. Continuous (aggregative) models describe the population using global statistics. The distribution of the genotypes and phenotypes in the population represents the individual's inherited behavior and the influence of the environment on the individual, respectively. Since the population's state is described as frequency data, the differences between individuals are lost in such a model. On the other hand, the discrete (agent-based) models maintain genotype/phenotype information of each individual in addition to other properties such as the location in the social network and spatial position (Alexander, 2010).

The fundamental difference between the two models is in computational complexity. Aggregative models can be expressed as a set of differential/difference equations, making it possible to find the solution analytically. On the other hand, solving problems solely using analytical techniques is not feasible with discrete (agent-based) models. Therefore, one must run a series of computer simulations and employ Monte Carlo methods to yield the solution (i.e., convergence behavior). However, despite being computationally less demanding and heavily utilized in solving multiplayer games, aggregative models cannot be utilized for modeling structured relations. Human interactions within society are represented as structured interactions, i.e., humans are constrained to the network of social relationships. That means interactions with close ones and their respective groups will significantly impact future behavior, unlike random strangers (Majhi et al., 2022; Szabó & Fáth, 2007). Therefore, utilizing aggregative models for modeling human interactions would be detrimental because structured interactions between individuals produce different outcomes compared to unstructured interactions (M. Nowak & May, 1992).

The introduction of the structure in evolutionary game-theoretic models dramatically influenced the model's long-term behavior (Alexander, 2010; Jusup et al., 2022). Embedding human-like social interaction structure into the structure of the agent-based models enables forecasting the less divergent long-term behavior, which resembles the actual human population. Therefore, such evolutionary game-theoretic models can account for a wide variety of human behaviors predicting the outcomes of many cooperative ethical dilemma games elaborated above, such as the Prisoner's dilemma, Stag Hunt, and fair division in the Nash bargaining game (Durrett & Levin, 1994; Jusup et al., 2022). It can be observed that, ultimately, the structure of society heavily influences the evolution of social norms (Alexander, 2010).

# 8 Ethics and coordination

If we put actors into (limited) material circumstances, we can expect that better-performing actors gain an advantage. In such circumstances, moral and ethical rules arise spontaneously to enable cooperation since greater coordinated groups are more effective than individuals if they have a similar developmental basis (Axelrod, 1984). It is argued that cooperation helped the human race survive in a discrepancy with competitiveness (Veit & Browning, 2020). Cooperation has been the basic organizational unit of the development of civilization since the time of hunter-gatherers (Diamond & Diamond, 1997).

Traffic is an excellent example of written and unwritten rules of conduct (Gintis, 2010; Helbing & Huberman, 1998). It is in the interest of every driver to cross the road from A to B as quickly and safely as possible. By refusing to follow the written rules, the driver risks being stopped by the police and losing his driving license (which in this case means expulsion from the game or losing the opportunity to participate). Failure to follow the unwritten rules carries the risk of condemnation, i.e., bad will by other players or their refusal to cooperate. Well-engineered traffic rules enable the transport system to work effectively and at increased performance.

The application of Prisoner's Dilemma is the same here. If the driver of the car *X* drives in an unknown place to which he will never return, selfish behavior, such as taking away advantage of and not letting other vehicles through, will bring him maximum short-term benefit. However, when the driver of car *X* does the same in the community where he is known, such behavior will bring him a bad reputation. Such stigma will negatively impact future rides regarding legal penalties and consequences outside the ride, e.g., degraded relations with community members. Whether a person has selfish or altruistic interests, both people know that in expectation, it is most profitable to follow the rules (Mueller, 1986). Violation of the rules can bring a one-time benefit, i.e., overtaking in the opposite lane over the full line, if the necessary conditions are met, and the person is not fined or physically punished for this procedure. If a person repeats this procedure, the chances of a positive outcome are reduced, and the person risks being excluded from traffic and being punished with some form of legal penalty, which means that in the long run, it is unprofitable to break the set rules consistently. In this case, it is opportune to follow the rules to get a satisfactory result, i.e., to reach the ride's goal. Legal codes of conduct in traffic are found on almost every part of the road regarding prohibitions, permits, or warnings making traffic an excellent example of legally enforceable and supervised ethics. Moreover, behavioral rules of individuals in society are another, yet more subtle, an example where ethical rules are unwritten. On the other hand, laws are an example of written applied ethics; however, it under-defines human interactions, further honed with unwritten (traditional, habitual) rules.

Another example of ethics (and law) are community standards and rules, for example, in online circumstances. Facebook uses agent-based models to simulate the effects of different rules (Ahlgren et al., 2020). Ethicists try out different rules and test for consequences in the system. This is a form of consequentialist exploration, whereby deontology is made based on rules' consequences (consequentially derived). Moreover, consequentialism relies on the principle of inherent cognitive limits unattainable to limited agents, especially in real-time. On the other hand, a simple set of rules is easy to follow, even in real-time, for a limited agent. Therefore, it makes sense to invest considerable effort in moral innovation to pre-calculate offline straightforward sets of rules that can be quickly followed under more strict limitations. The principle of offline pre-calculation of ethical rules in conditions with enough time and computation resources is similar to planning and acting under time constraints.

# 9 Human values and coordination

In addition to moral and legal obligations, there is also the issue of human values. Coordination is non-trivial, even hard to achieve. Mathematical-computational models and their analysis can reinforce the previous statement (Aaronson, 2013; Daskalakis et al., 2006). For example, the problem of finding Nash equilibrium is PPAD-complete; hence solving it might take prohibitively long. This is the case both in single (Daskalakis et al., 2006) and iterated settings of problems (Borgs et al., 2008), despite the folk theorem and abundance of Nash equilibria in the latter. Additionally, Nash equilibrium is achieved by rational actors only if they share beliefs about how the game is played. The rational actor model has no inherent mechanisms to enforce shared beliefs, so complex Nash equilibria do not arise spontaneously between rational agents. On the other hand, there is a concept of *correlated equilibrium* that is an appropriate equilibrium concept for social theory (Gintis, 2009). It is a generalization of Nash equilibrium which includes the correlation device in the model that induces correlated beliefs between the agents. Correlation devices can take the form of shared playing history, selection of players, public signals (like group symbols), etc. (Morsky & Akçay, 2019). Additionally, finding a correlated equilibrium is much easier than Nash equilibrium as it can be done in polynomial time for any number of players and strategies in a broad class of games by using linear programming, even though finding the optimal one is still NP-hard (Papadimitriou & Roughgarden, 2008).

Values are legally, and morally undefined items individuals elevate, value, and cultivate because of cultural and personal prejudices (Mead, 1976). We hypothesize that shared values ingrained in us through culture are an emerging phenomenon that helps with coordination in a fast heuristic fashion. This is in line with results suggesting that morality judgements are driven at least partly by imprecise heuristics and emotions (Capraro & Rand, 2018). Capraro & Perc, 2021 have mathematically modeled moral preferences by augmenting single-objective utility function with a weighted (scalarized) term for following personal norms (in addition to monetary outcomes).

*Human values* are an emerging concept that allows for easier coordination among like-minded people within a community. Suppose a person makes judgments based on a pre-judgment created by the human values defined above. There is an increased chance that the foundation will lead the person to a different conclusion from someone with different human values. If we have correlated values, we have a similar basis for decision-making, hence heuristically aiming for a correlated equilibrium.

Norms as sets of rules and conventions could also be a correlating device if they are simple enough to follow (Axelrod, 1984). They should at least be explainable and comprehensible (Dosilovic et al., 2018; Juric et al., 2020; Krajna, Brcic, Kovac, et al., 2022; Krajna, Brcic, Lipic, et al., 2022). However, trying to follow many rules is certainly computationally hard, as constraint satisfaction problems from computer science can attest (Gallardo et al., 2009). Using continuous fields of values enables using approximate-continuous instead of combinatorial reasoning, making it a very effective mechanism that can be seen in today's deep neural networks. If we were to use combinatorial reasoning in complex and fast situations, we would be paralyzed in decision-making under our cognitive limits, and coordination would be rare (Talbert, 2017). Even worse would be trying to calculate Nash equilibrium on the fly, outside the realm of games with a choreographer.

From a philosophical and psychological point of view, human values can be represented as a mixture of clustered criteria individuals use to evaluate actions, people, and events. Moreover, the Values Theory identifies ten distinct value orientations common among people in all cultures. Those values are derived from the three universal requirements of the human condition: biological needs of individuals, requisites of coordinated social interaction, and groups' survival and welfare needs (Schwartz, 2006). Individuals communicate these ten values with the remainders of the group in order to pursue their goals. According to the Values Theory, these goals are described as trans-situational and of varying importance serving as guiding principles in people's lives (Schwartz, 1997). Figure 1 depicts the ten values in a circular arrangement so that the distance and antagonism of their underlying motivation are inversely proportional, i.e., two close values share a similar motivation, and two opposite values have opposing motivations. Moreover, values can be divided into two planes: *self-enhancement* (pursuit of self-interest) versus *self-transcendence* (concern for the interests of others) and *openness* (independence and openness to new experiences) versus *conservation* (resistance to change).

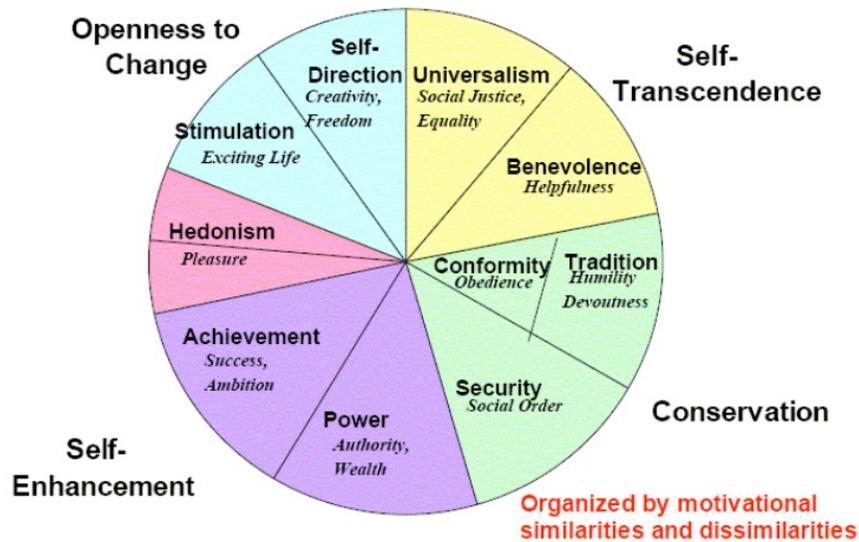

*Figure 1: Ten motivational values and their relationship (Schwartz, 2006)*

# 10  Evolutionary game theory (EGT) as modeling basis

A commonly accepted hypothesis is that evolutionary processes shaped life on earth. Similarly, EGT can be used to determine the influence of external forces on heterogeneous ethics as its underlying component. Therefore, a possible explanation of those external forces is the necessity for cooperation as the basis of heterogeneous ethics on which all today's civilizations are built. It is argued that EGT is a realistic explanation of the material circumstances that preceded the creation of the first unwritten moral rules of cooperation (Cavagnetto & Gahir, 2014; O'Connor, 2020).

EGT implements the three main pillars: 1) higher payoff strategies over time replace lower payoff strategies, also known as the "survival of the fittest"; 2) evolutionary change does not happen rapidly; 3) players' future actions are made reflexively without reasoning (Friedman, 1998). Biologists and mathematicians initially developed the evolutionary game to resolve open questions in evolutionary biology (Smith, 1982). However, it has far-reaching implications in many areas such as economics, ethics, industrial organization, policy analysis, and law. In general, evolutionary game models are suitable for systems where agents' behavior changes over time and interacts with the other agents influencing their behavior. However, the other agents must not collectively influence the behavior of an individual agent, and all decisions must be made reflexively (Friedman, 1998). One downside to original EGT is that players (agents) are born with a particular strategy that cannot be changed during their lifespan (Dugatkin & Reeve, 2000), making it unrealistic to model humans who evolve and change their strategy of social interactions throughout time. However, EGT, as it is currently defined, is suitable for modeling reptiles that do not have strong learning capabilities (Matsubara et al., 2017). Simple, parametric forms of learning through memory and reputation-mechanisms have been implemented in EGT, but it does not include richer lifetime learning due to computational complexity concerns.  For modeling humans, extensions of EGT should be investigated to find concepts interpolated between stable evolutionary strategies and Nash equilibria since the first is reactive (without deliberation) while the second is unrealistically rational and computationally demanding. Correlated equilibria seem good direction that possesses good balance of power and efficiency.

The usage of EGT as a basis for modeling morality has been extensively discussed in previously mentioned by Alexander in his *The Structural Evolution of Morality*, where he recognized the deficiencies of solely using EGT and proposed utilizing the combination of EGT, theory of bounded rationality, and research in psychology (Alexander, 2010). Although EGT enables the identification of behavior that maximizes the expected long-term utility, the motivation behind this behavior and the subsequent action that complies with the moral theory remains unexplained. For an individual to maximize his lifetime utility, one's actions must be bounded by rationality, requiring reliance on moral heuristics such as fair split and cooperation. Consequently, incorporating bounded rationality into one's actions complies with moral theory (Veit, 2019).

Authors in **extended evolutionary synthesis** propose an improvement on systems focused on genetic evolution by considering co-evolution of genome and culture. Cultural evolution alters the environment faced by genes, indirectly influencing natural selection. Adding social norms with the possibility of arbitration can substantially widen the range of successful cooperation (Mathew et al., 2013). This can explain the ultra-sociality of the human species. This co-evolution supposedly creates multiple equilibria, among which many are group-beneficial.

According to this line of thinking, in-group competition solves the free-rider problem with punishments, reputation, and signaling which are mechanisms for large-scale cooperation. It sustains adherence to norms and it settles the group into some correlated equilibrium. What is special about these mechanisms is that they can sustain any costly behavior with or without communal benefit. That is, they can sustain social norms that need not necessarily be cooperative norms (Henrich & Muthukrishna, 2021).

Cultural evolution is a much faster and more innovative information processing system. Unlike genetic evolution where there are two models for recombing traits, here there are many more models simultaneously from which cultural traits are interacting. Additionally, transmission fidelity is much lower and selection is strong and influenced by psychological processes which drives greater innovation (Henrich & Muthukrishna, 2021). As it is known, the success of strategies in a population is conditional on the populational distribution of other strategies and these conditions can shift fast in changing. Using cultural learning, individuals can quickly adapt behavior to circumstances for which genetic learning is too slow by imitation learning and can keep cooperation from collapsing. Hence, culture may have created prolonged cooperation based on indirect reciprocity which may have been just enough for genetic evolution to pick it up to develop supportive psychology to perpetuate it.

Cultural evolution is more likely to create inter-group competition since it is fast, noisy, and nonvertical compared to genetic evolution (Boyd et al., 2011). This competition puts groups against each other performance-wise and it tends to lead to more prosocial norms and institutions. Competition at a lower level (of smaller groups) can help cooperation at higher levels (of greater collectives) and vice versa, stronger cooperation at a lower level can be detrimental to cooperation at a higher level (Muthukrishna et al., 2018). Sometimes inter-group competition leads to weakening kin bonds, reducing effectiveness at lower scales to promote effectiveness at higher scales (Henrich & Muthukrishna, 2021).

EGT is rather successful in modeling social phenomena due to interactions between individuals trying to maximize their utility. The emergence of altruism in an *n*-player prisoner's dilemma using EGT is proposed by (Fletcher & Zwick, 2007). Authors suggest that utilizing an EGT approach has shown to be useful in understanding the inherited similarities between weak and strong altruism. The influence of social learning on human adaptability is discussed in (Kameda, 2003). By using the EGT approach to model the social learning of individuals through selective imitation, the authors supported the hypothesis. The development of social norms as an evolutionary process is another example of modeling social phenomena. Evolutionary psychologists argue that humans are not skilled at logical problem-solving (Clark & Karmiloff-Smith, 1993). Therefore, humans do not reason what is true or false when faced with reasoning; they match different patterns to a particular case. In (Baimel et al., 2021) it was shown that human development is more consistent with cumulative cultural learners than with Machiavellian intelligence that tries to strategically outmaneuver an opponent. People will use previously learned reasoning that includes actions of what is obligated, permitted, or forbidden. Such reasoning is inherited by social norms and, consequently, can be justified using EGT (Ostrom, 2000). (Bicchieri, 2005) describes how social norms, through sanctions, transform mixed-incentive games with social dilemmas where cooperative outcomes are unstable into easier coordination problems.

ESS conditioned on cues from public signals have been proved to be correlated equilibria of the game (Cripps, 1991) and these equilibria can be found by repeated play (Arifovic et al., 2019). (Gintis, 2010; Morsky & Akçay, 2019) model social norms that act as "choreographers" that induce correlated beliefs in agents, allowing them to coordinate on a correlated equilibrium of the game.

EGT and its extension to genetic-cultural co-evolution can model dynamics, progress as well as limits. What is necessary is to incorporate cognition and more complex learning and strategies into cultural processes to make more precise dynamic models of changes. Additionally, a mixture of games should be modeled on a set of players, with uncertainty surrounding the specifics of the game played and outcomes. Such players would have evolving interests that depend on a selection model that mirrors the one in humans. Something along that line of thinking, but outside of ethical considerations, was done in machine learning for solving a considerable set of tasks with the same agent (Reed et al., 2022). In social physics, some progress has been achieved in multigames (Li et al., 2021) and in modeling more complex group dynamics with higher-order interactions (Majhi et al., 2022).

## 11  Social norms as behavioral patterns

Ethics consist of behavioral patterns/regularities (social conventions, of which norms are a subset) that can be observed in resolutions of recurring coordination problems-situations (Bicchieri, 2005; Lewis, 1969) *in a society of agents with similar capabilities*. These patterns are emergent through time from the interactions in the environment. Under the assumption of evolutionary-guided changes (e.g., genetic-cultural co-evolution), all circumstances that often appeared in time were used for selective pressure (Henrich & Muthukrishna, 2021). For these reasons, it is expected that such norms would be locally consistent and good performing in frequent circumstances that led to their creation. (Morsky & Akçay, 2019) have shown that natural selection can serve as a blind choreographer that spontaneously creates beliefs and norms from stochastic events so that they can serve as correlated equilibria without sophisticated knowledge or external enforcement. These beliefs and norms can be sustained using initially simpler and later more complex mechanisms (Bhui et al., 2019).

During their lifetime, humans face various situations where certain decisions must be made. Such situations are simply a part of life, and we cannot avoid them. However, making decisions and confronting the resulting consequences is under our power. Throughout the evolution of humankind, individuals have been confronted with various decisions passed to and replicated by others over generations. Over time, the aggregation of these decisions led to the development of ethics, which can be compared to a patchwork, as depicted in Figure 2. Every patch in patchwork represents a similar set of situations (episodic games) and belonging norms. However, the two neighboring patches are similar in problem space but have different norms that govern them. Also, the white patches represent the absence of norms in certain areas due to the absence of lived experience in that space. Examples of such white patches might be situations involving significantly novel and impactful technology (such as super-intelligence). In that case, humans have to extrapolate norms from neighboring patches, i.e., similar ethical settings. The extrapolation, if it may be simply and uniquely done in the first place, is not guaranteed good performance or relevance.

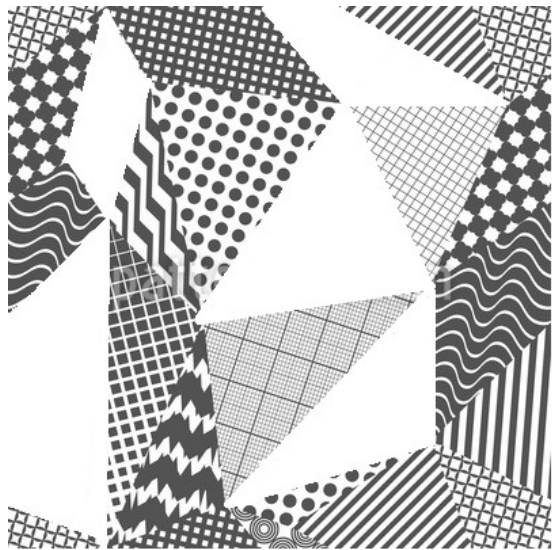
*Figure 2: Ethical system - patchwork of norms*

# 12  Consistency of ethics

In addition to multicriteriality as a source of dilemmas for all normative ethics, deontological ethical systems may additionally experience dilemmas through inconsistencies. Two mechanisms yield inconsistency: norm confusion and faulty execution.

When extrapolating to substantially new situations from existing patterns, we might get **norm confusion -** inconsistencies between the different patches of locally consistent patterns. It is unclear which norm should be applied, and we get into a dilemma (Bicchieri, 2005). These inconsistencies are problematic for the algorithmization and alignment with future AI systems.

All existing ethics contain inconsistencies, with evident contradictions if looked at from a high-enough level. Such inconsistencies are not necessarily visible locally.

**Faulty execution** yields moral inconsistency in humans based purely on emotions associated with a particular case, i.e., moral dilemma, and not on formal inconsistencies. The root cause of such inconsistencies occurs when an individual, faced with dilemmas, treats the same moral cases differently. The process of learning from such previously made mistakes and the process of self-improvement allowing the individual to maintain consistency with moral norms shared within society is called *moral learning*. However, avoiding moral inconsistencies through moral learning is not always straightforward due to conflict with self-interest. Moreover, moral norms are generic, i.e., they are applied to a wide array of cases, and, consequently, there will always be exceptions. For an individual (learning agent) to learn through moral problems on their own, one must think about moral problems from the other's perspective. For example, using Bayesian reasoning one can derive a clear moral rule based on the judgments of other individuals (Campbell, 2017).

To avoid biases when dealing with ethical decisions, philosopher John Rawls proposed the *Veil of Ignorance* as a tool for increasing personal consistency regarding some forms of faulty execution. Here, one should imagine sitting behind a veil of ignorance, keeping him away from his identity and personal circumstances. By being ignorant in such a manner, one can objectively make decisions. This would lead to a society that should help those who are socially or economically lacking behind (Rawls, 1971) because robust optimization under total ignorance yields a maximin solution.

# 13  Cooperation vs competition

The question of the place of competition within well-functioning societies is open for investigation. It is argued that ethics based on cooperation brings greater progress in the long run (Cliquet & Avramov, 2018; Curry, 2016). However, the relationship and the balance between the two is complex, even if the desired final goal is worldwide cooperation. Social physics exhibits a complex relationship between the emergence of the two that is very sensitive to the setting of the problem at hand. Competition in society seems to play both an **innovative** and **cohesive role** in cooperation.

In addition to being the basis for the development of civilization, cooperation incorporates individual and social interests and helps create a balance among community members. On the micro-level, in civilized societies and everyday activities, cooperation with other community members in the long run is more profitable due to the installed norms and institutions. This makes personal goals faster and easier while achieving greater communal well-being. Opposite to cooperation is competitiveness, which in itself is not bad. Competition seems to be one of the drivers of innovation, while cooperation is more effective at operational issues in repeated situations. It is good to be competitive with, for example, a past version of ourselves, set personal goals, and fight to achieve them. Additionally, the competition seems to be a cohesive element of cooperation.

According to the extended evolutionary synthesis, **in-group competition** is important to solve the free-rider problem through mechanisms of punishment, reputation, and signaling. Hence, it improves adherence to the group norms. This efficiently leads to the correlated equilibrium.

**Inter-group competition** is important solely for correlated equilibrium selection, i.e. search. In line with theoretical results, searching for optimal correlative equilibrium is a painfully slow process and it can be incomplete to remove group-damaging norms – especially when the latter are entangled with important cooperative norms. Also, the balance and distribution of competition and cooperation are sensitive, whereby competition on lower levels can favor cooperation at higher and stricter cooperation on lower levels can lead to collapse on a higher level. Such complex group dynamics can be modeled and tested on graphs (Szabó & Fáth, 2007) and hypergraphs (Majhi et al., 2022).

In the long term, cooperation outweighs competition when relying on scarce resources. This is best described in Hardin's *The Tragedy of the Commons* (Hardin, 1968)*,* where each individual consumes resources at the expense of the others in a rivalrous fashion. If everyone acted solely upon their self-interest, the result would be a depletion of the common resources to everyone's detriment. The solution to the posed problem is the introduction of regulations by a higher authority or collective agreement, which leads to the correlated equilibrium (Ostrom et al., 1994). Regulations could directly control the resource pool by excluding the individuals who excessively consume the resources or regulating consumption use. On the other hand, self-organized cooperative arrangements among individuals can rapidly overcome the problem (with a punishment mechanism for deviators). Here, the individuals share a common sense of collectivism, making their interest not to deplete all resources selfishly (Hardin, 1968).

# 14  Conclusion

We have looked at ethics through an analytical prism to find some of its constitutive properties. The problem that ethics tries to solve is multi-criteria, dynamic, and poised by uncertainties. We argue that ethics is related to multi-agent interaction so it can be properly modeled by game theory. Ethical systems are group-performance oriented, have collateral origins, and consist of honed behavioral patterns – social norms. Ethical norms can be approximated by values. Current ethical systems are globally inconsistent, though they are locally consistent. Ethics is focused on cooperation, but it also depends on the competition for efficiency and adaptability. Moreover, the balance between competition and cooperation is delicate. The levels at which competition takes place greatly impact the level at which beneficial cooperation emerges, if at all. Mechanisms such as reputation, signaling, and punishment are elements of in-group competition that drive group cohesion. Social norms with in-group competition play a crucial role of a correlation device that enables finding a correlated equilibrium into which a group may settle in a computationally efficient manner. This is in stark contrast to the problem of finding Nash equilibrium which is PPAD-complete, and solving it might take a prohibitively long time. However, there are no guarantees that the found correlated equilibrium is beneficial to its group. Inter-group competition drives a slow search for better equilibria. This slowness is in line with the results from the computational complexity theory. The optimal ethical system could be computationally found in principle, though at an impractically great computational cost.

Brcic & Yampolskiy, 2021 argue that ethics should stop being collateral and that through modeling we can take more control over the process of ethical developments to obtain codifiable, more consistent, and adaptable ethics. The aim would be to make alignment easier within vast aggregates of agents (humanity, AI, inforgs (Floridi, 1999)) through co-evolution between different constituents that is guided by a set of meta-principles. We propose agent-based models of genetic and cultural co-evolution in a similar way to (Paolucci et al., 2012). However, agents in these models should also be equipped and amplified with basic cognition, and more complex learning and strategies as idealizations to which we may strive in simulations according to available computational and algorithmic resources. This considerable computational effort can be invested in moral innovation to pre-calculate offline straightforward sets of rules, norms, and values that can be quickly and stably followed under more strict real-time limitations. More generally, findings from this research direction might benefit decentralized computing on an unprecedented scale and heterogeneity. If any of this is possible and practical, it remains to be seen.

# Acknowledgments

We thank Jeroen van den Hoven (Delft University of Technology) for useful comments and remarks regarding existing research in the field that increased our coverage.